\documentclass{aa}
\usepackage{graphicx}
\usepackage[latin1]{inputenc}
\usepackage{txfonts}
\usepackage{natbib}

\begin{document}

\title{Galaxy properties from voids to clusters in the SDSS-DR4}
\titlerunning{Galaxy properties from voids to clusters}

\author{G. Sorrentino \inst{1,3,4},  
	V. Antonuccio-Delogu \inst 2 
	\and
	A. Rifatto \inst{1}
	} 
\authorrunning {Sorrentino, Antonuccio-Delogu \& Rifatto}

\institute{
INAF-Osservatorio Astronomico di Capodimonte, Via Moiariello 16, 80131 Napoli,
ITALY
\and
INAF-Osservatorio Astrofisico di Catania, Via S.Sofia 78, 95123 Catania, ITALY
\and
Dipartimento di Fisica e Astronomia, Universit\'{a} di Catania,
Via S.Sofia 78, 95123 Catania, ITALY
\and
INAF-$VSTceN$, via Moiariello 16, 80131 Napoli, ITALY}

\abstract
{}
{We investigate the environmental dependence of galaxy population properties
in a complete volume-limited sample of 91566 galaxies in the redshift range
$0.05 \le z \le 0.095$ and with $M_r \le -20.0$ (that is $M^* + 1.45$), selected
from the Sloan  Digital Sky Survey (SDSS) Data Release 4 (DR4). Our aim is to
search for systematic variations in the properties of galaxies with the local
galaxy density. In particular, we analize how the ($u - r$) color index and the
morphological type of galaxies (the latter evaluated through the SDSS {\em
Eclass} and {\em FracDev} parameters) are related to the environment and to the
luminosity of galaxies, in order to find hints that can be related to the
presence of a ''void'' galaxy population.}
{"Void" galaxies are identified through a highly selective criterion, which takes
also into account redshift and allows us to exclude from the sample all the
galaxies that are not really close to the center of underdense regions. We
study the ($u - r$) color distribution for galaxies in different luminosity bins,
and we look for correlations of color with the environment, the luminosity, and
the morphological type of the galaxies.}
{We find that galaxies in underdense regions (voids) have lower luminosity
($M_r>-21$) and are bluer than cluster galaxies. The transition from overdense
to underdense environments is smooth, the fraction of late-type
galaxies decreases while the fraction of early-type galaxies increases smoothly
from underdense to dense environments.}
{We do not find any sudden transition in the galaxy properties with density, which, according to a suggestion by Peebles (2001), should mark the transition to a population of "void" galaxies in  
$\Lambda$CDM models. On the contrary, our results suggest a continuity of 
galaxy properties, from voids to clusters.}

\maketitle

\section{Introduction}

A self-consistent theory of galaxy formation and evolution within the
gravitational collapse paradigm has yet to come, although quite significant
advancements have been recently done (e.g. \citet{2004AIPC..743...33S} for an
up-to-date review). Such a theory should explain the observed properties of
galaxy populations in very different environments, ranging from very dense
clusters to large voids. However, the mere existence of galaxies within voids
is a challenge for any gravitational bottom-up scenario of the Large Scale
Structure (LSS) formation in the Universe, because in this scenario voids form
out of the expansion of underdense ($\delta < 0$) initial density fluctuations.
Theoretical models and numerical simulations
\citep[e.g.][]{1979ApJ...233..395O,2005MNRAS.362..213B} suggest that
dark matter flows out of voids, and partly concentrates in some thin
boundary shells. The latter, however, have not been detected,
  although there is clear evidence that the galaxy density profiles
  inside voids become steeper near their borders
  \citep{1996A&A...314....1L,2004ApJ...607..751H,2005ApJ...621..643G}. 
  Observationally, there is evidence that voids are surrounded by superclusters of
  galaxies, as is vividly shown by recent analyses of the Las Campanas
  and SDSS surveys \citep{2003A&A...410..425E, 2003A&A...405..425E, 2005A&A...439...45E}. 
  It is interesing to note that the void models of \citet{1979ApJ...233..395O} and 
  \citet{2005AAS...206.1002H} are only concerned with the global
dynamics of dark matter (DM): much less work has been devoted to
understand the nonlinear collapse of DM halos and the formation of
galaxies {\em within} voids.  \citet{1986ApJ...303...39D} suggested
that voids could be populated by a population of faint dwarf galaxies,
where star formation is suppressed by strong photoionization occurred
at $z > 6$ \citep{2005astro.ph..1304H}.\\
\noindent
The search for galaxies in voids started with the discovery
  of the voids themselves (i.e., \citet{1996A&A...314....1L} for a full
  review in this field until 1996). \citet{1996A&AS..116...43P,
  1997A&A...325..881P} looked for emission-line galaxies in
  nearby voids, concluding that most of them are found close to the
  borders, and far from the deep internal regions.
There have also been
(failed) attempts at detecting a population of dwarf galaxies more
homogeneously distributed within voids \citep{2003MNRAS.341..981S,2005MNRAS.357..819S,
2004MNRAS.352..478R}, but the lack of an evident faint end of the
dwarf galaxies luminosity function does not allow to make firm
quantitative predictions about their abundance.\\
Since the seminal papers by \citet{1999AJ....118.2561G,2000AJ....119...32G},
some efforts have been dedicated to detect and to study the properties of
bright galaxies in nearby voids. The availability of large samples of galaxies
from surveys like the 2dF and the Sloan Digital Sky Survey (SDSS) has made
possible more complete analysis of the dependence of galaxy properties on the
environment. \citet{2005MNRAS.356.1155C} have shown that the luminosity
function in voids tends to have a steeper slope at high luminosity, but not in
the fainter range. The dependence of ($u - r$) color distribution on the
environment has also been recently studied by \citet{2004ApJ...615L.101B}, who
found that its shape is bimodal and can be well described by a linear
combination of two gaussian distributions for a very wide range of local
densities. Surprisingly, they also find a small quantity of {\em red},
early-type galaxies even in the lowest density environments, thus suggesting
that there is not a rigid morphological or color segregation with environment.
However, the environmental dependence of galaxy properties is only detected for
galaxies in the luminosity range $-20.4  < M_{r} < -19.4$
\citep{2004AJ....128.2677T}: for galaxies brighter than $M_{r} = -20.4$ there
are very little variations of their properties with the environment.
\citet{2001ApJ...557..495P} suggested that galaxies lying within very
underdense regions are representative of a distinct population. In this case,
we should observe a transition in one or more galaxy properties moving from
clusters to voids.\\
In order to verify the previous hypothesis, we will investigate the variation
of the {\em color distribution}, {\em morphology}, and {\em luminosity} with
the local density. This kind of analysis has been made possible by the
availability of a large sample of galaxies extracted from the SDSS - Data
Release 4 
\citep[SDSS-DR4:][]{2000AJ....120.1579Y,2002AJ....123..485S,2004AJ....128..502A}.
Using this huge data sample we confirm some previous results
\citep[e.g.][]{2004ApJ...617...50R}, and we are able to extract statistically
significant subsamples for different relative underdensities. We will study the
correlation between ($u - r$) colors and the environment for {\em bright} and
{\em faint} galaxies, also looking for sudden changes which should mark the
emergence of a distinct void galaxy populations, as suggested by Peebles
(2001). We will use a criterion to estimate the density which is different from
the $\Sigma_{5}$ criterion used by, e.g, \citet{2004AJ....128.2677T}; on the
contrary, it is more similar to the one adopted by \citet{2004ApJ...617...50R},
because we want to be sure that galaxies in voids are correctly identified and
included in the sample.\\
The plan of the paper is as follows: after the description of the galaxy sample
(sect. 2), in section 3 we will discuss the adopted criteria to construct our
subsamples. In section 4 we will present our results concerning the ($u - r$)
color distribution, and its dependence on the local density and luminosity. In
section 5 we will analize the results of the previous section taking also into
account the relationship between the ($u - r$) color distribution and the
morphological type of galaxies, then discussing the relationship with the
morphology-density relation (Dressler 1980). Finally, in section 6
and in section 7 we will present a general discussion and our conclusions.\\
Throughout, we adopt standard present day values of the cosmological parameters
to compute comoving distances from redshift: a density parameter $\Omega_m =
0.3$, a cosmological constant $\Omega_{\Lambda} = 0.7$, and a Hubble`s constant
value $H_0 = 75~km~s^{-1} Mpc^{-1}$.
\section{SDSS - DR4}
The SDSS
\citep{2000AJ....120.1579Y,2004AJ....128..502A,2005AJ....129.1755A,2005astro.ph..7711A}
is a photometric and spectroscopic survey, which will map about one quarter of
the entire sky outside the Galactic plane, and will collect spectra of  about
$10^{6}$ galaxies, $10^{5}$ quasars, 30,000 stars and 30,000  serendipity
targets. \\
Photometry is available in {\itshape $u'$, $g'$, $r'$, $i'$} and {\itshape
$z'$} bands \citep{1996AJ....111.1748F,1998AJ....116.3040G},  while the
spectroscopic data are obtained with a pair of multi-fiber spectrographs.  Each
fiber has a diameter of 0.2 mm (3$\arcsec$ on the sky), and adjacent fibers
cannot be located more closely than 55" on the sky ($\sim$ 110 kpc at $z$ = 0.1
with H$_0$ = 75 km s$^{-1}$ Mpc$^{-1}$) during the same observation. In order
to optimize the placement of fibers on individual plates, and as the placement
of plates relative to each other, a tiling method has been developed which
allows a sampling rate of more than 92\% for all targets. For details see the
SDSS web site (www.sdss.org/dr4/algorithms/tiling.html).\\
The spectroscopic SDSS-DR4 catalog contains about 673,280 galaxies and  covers
an area of 4783 square degrees. The spectra cover the spectral range
$3800<\lambda<9200$ \AA, with a resolution of
$1800<\lambda/\Delta\lambda<2100$, and give a rms redshift  accuracy of $30$
Km~s$^{-1}$, to an apparent magnitude limit (Petrosian magnitude) of
$r'=17.77$.\\ Data have been obtained from the SDSS database
(http://www.sdss.org/DR4) using the CasJobs facility
(http://casjobs.sdss.org/casjobs/).
\section{Sample selection}
\begin{figure}[htbp]
\begin{center}
\includegraphics[width=8cm]{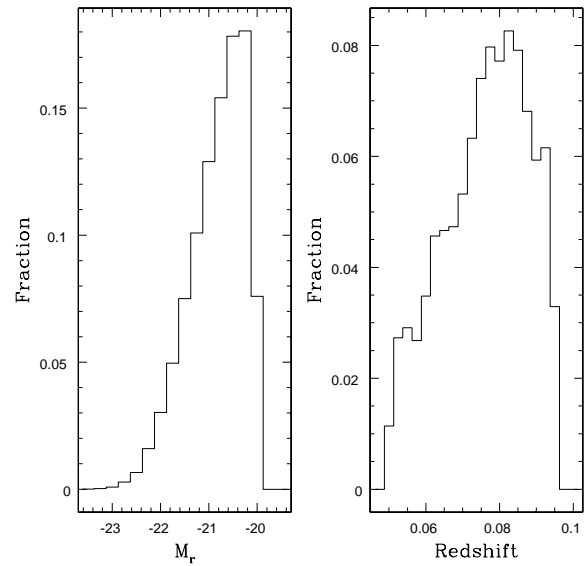}
\caption{The redshift (right), and r-band absolute magnitude distributions
(left), for the 91566 galaxies in our sample. The absolute magnitude
corresponding to the upper redshift limit ($z = 0.095$) is $M_{r}=-20.0$.}
\label{fig:z}
\end{center}
\end{figure}

In this paper, we will take into account a complete volume-limited sample of
galaxies in the redshift range $0.05 \le z \le 0.095$, brighter than 
$M_r =-20.0$, that is $M_r^*$ + 1.45, with $M_r^* = -21.45$. 
The lower redshift limit is chosen with the aim of  minimizing the aperture bias
\citep[e.g.][]{2003ApJ...584..210G} caused by the presence of large nearby
galaxies, while the upper limit is aimed at obtaining a high level of
completeness, estimated through Schmidt's $V/V_{max}$ test.\\ 
Our initial sample contains 91566 galaxies. For each galaxy, we compute its
r-band absolute magnitude, dereddened and K-corrected as suggested in 
\citet{2003ApJ...592..819B}. The distributions of $z$ and $M_{r}$ are given in
Fig.~\ref{fig:z}.\\
Galaxies in our sample with no detectable emission lines in their spectra are
classified as {\em Passive Galaxies} (PGs): for these galaxies, the
morphological type should be earlier than Sa. Galaxies with one or more
emission lines having  $I_{\lambda}/\sigma_{I_{\lambda}} > 2$, where
$I_{\lambda}$ is the emission line flux and $\sigma_{I_{\lambda}}$ its
uncertainty, are classified as star forming galaxies (SFGs), according to the
criteria adopted by \citet{2001ApJ...556..121K}: for these galaxies, the
morphological type should be later than S0/Sa. In order to avoid all the
ambiguous cases in the AGN/SFG classification, we removed those sources whose
line ratios fall close to the border line of the diagnostic diagrams. This was
done by keeping only  those sources for which part of the error bar associated
to the logarithm  of the line-ratios lie within the theoretical uncertainty of
the model ($\sigma_{mod}=0.1$ dex) in both $x$ and $y$ directions.\\
Moreover, we adopted another criterion to separate the galaxies in early- and
late-type, according to their morphological type defined by the two parameters
{\sc eclass} and {\sc fracDev} provided by the SDSS. {\sc fracDev} is a
photometric parameter providing the weight of a deVaucouleurs component in best
composite exponential+deVaucouleurs models, and {\sc eclass} is a spectroscopic
parameter giving the spectral type from a principal component analysis.
Early-type galaxies (E + S0) were selected following the criteria adopted by
Bernardi et al. (2005): {\sc fracDev}(r) $> 0.8$ and {\sc eclass} $<0$.
Late-type galaxies (Sa and later) were selected when either {\sc eclass} $\ge
0$ or {\sc fracDev}(r) $<0.5$. In this way we exclude from our analysis all the
galaxies for which an unambiguous classification is not possible, because
their {\sc eclass} and {\sc fracDev} parameters are out of the previous
defined ranges.\\
In order to investigate a possible dependence of galaxy colors on the
environment, for each galaxy we compute the number of neighbours within a fixed
radius $D_{\rm max} = 5$ Mpc, the distribution and the related percentile
(Tab.~\ref{tab:percentile}). We adopted this particular distance because it was
used in previous works aimed at studying galaxies in very underdense regions
(voids) \citep{1999AJ....118.2561G, 2004ApJ...607..751H}. The use of the
percentile in our analysis allows us to compare different environments and to
characterize the difference between over- and underdense regions without 
defining an environment {\it a priori}.\\
A galaxy {\it j} is considered as a neighbour of a galaxy {\it i}
if:
\begin{itemize}
	\item $D_{ij} \le D_{\rm max}$     
	\item $c|z_i - z_j| \le 1000$ km s$^{-1}$ 
\end{itemize}
\noindent
where $D_{ij}$ is the projected distance between the two galaxies, and $|z_i -
z_j|$ is their redshift difference. $D_{ij}$ is computed from the angular
separation $\theta_{ij}$ and the redshift $z_i$; the limit $c |z_i - z_j| \le
1000 $\ km \ s$^{-1}$ is the value usually adopted to select cluster or galaxy
group members in the velocity space \citep{1996ApJ...473..670F,
2005MNRAS.358...88W}. Throughout, we intend as {\it local galaxy density} the
number of neighbours brighter than $M_r = -20.00$ and within $5$ Mpc.\\
In Appendix A we explore the effect of varying our definition of {\it local
galaxy density} on scales $2.5$ Mpc to $10$ Mpc. We find that our conclusions
remain unchanged.\\

\begin{table}[htbp]
\begin{center}
\caption{Numbers of neighbours and the related percentile. A percentile is
defined as the value of a given scalar quantity characterizing the fraction of
the distribution that is equal or smaller than that value.}
\label{tab:percentile}
\begin{tabular}{||r|r||}
\hline
 $N_{neigh}$           & Percentile \\
\hline
\hline

 0 		       &$ 1st$  \\
\hline
 2 		       &$ 5th$  \\
\hline
 4 		       &$10th$  \\
\hline
 9 		       &$25th$  \\
\hline
 18		       &$50th$  \\
\hline
 35		       &$75th$  \\
\hline
 59		       &$90th$  \\
\hline
 79		       &$95th$  \\
\hline
 127		       &$99th$ \\
\hline
\end{tabular}
\end{center}
\end{table}
\section{($u - r$) color distribution: dependence on environment and luminosity}
\begin{figure*}[htbp]
\begin{center}
\includegraphics[width=14cm]{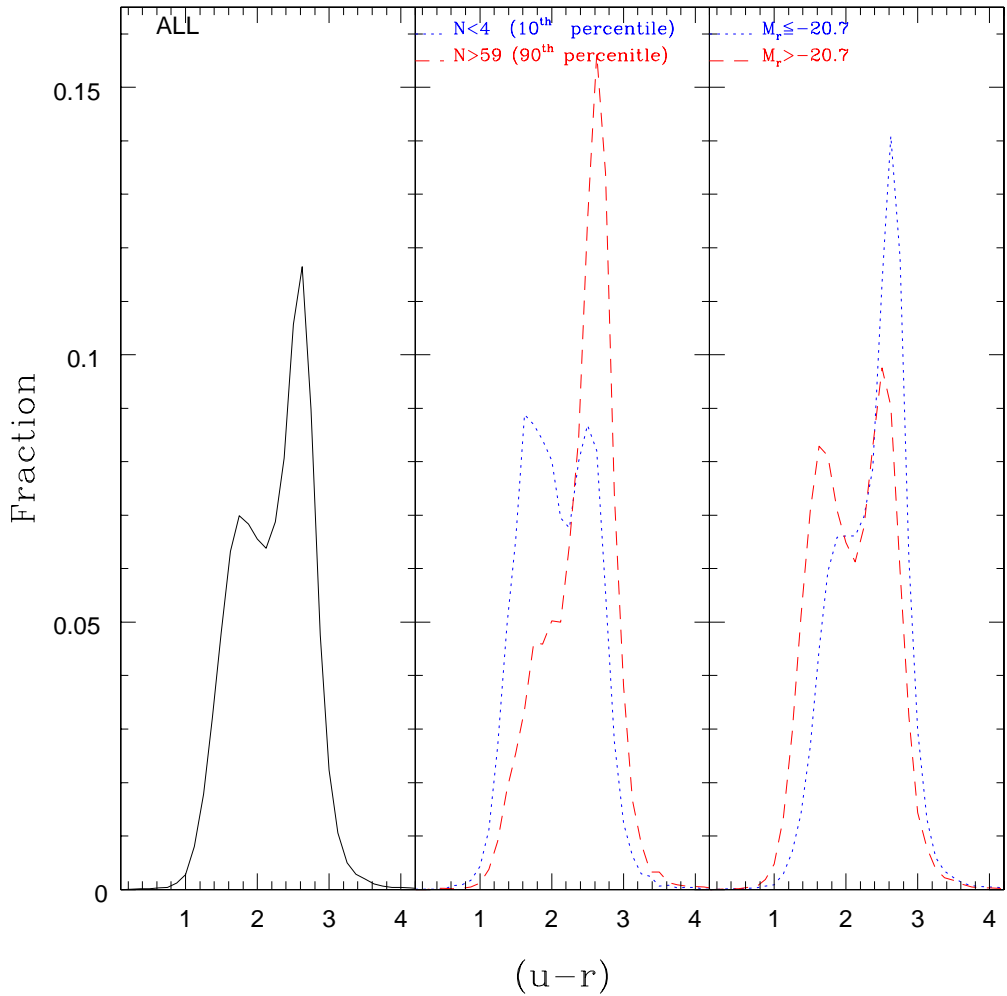}
\caption{Left panel: the ($u - r$) color distribution for all the galaxies in our
sample. Central panel: the ($u - r$) color distribution for galaxies  having less
than 4 neighbours (corresponding to 10th percentile) compared with the
distribution of galaxies having more than 59 neighbours (corresponding to 90th
percentile) within 5 Mpc. Right panel: the ($u - r$) color distribution for
galaxies brighter and fainter than $M_r = -20.7$.}
\label{fig:ur1}
\end{center}
\end{figure*}
\begin{figure*}[htbp]
\begin{center}
\includegraphics[width=12cm]{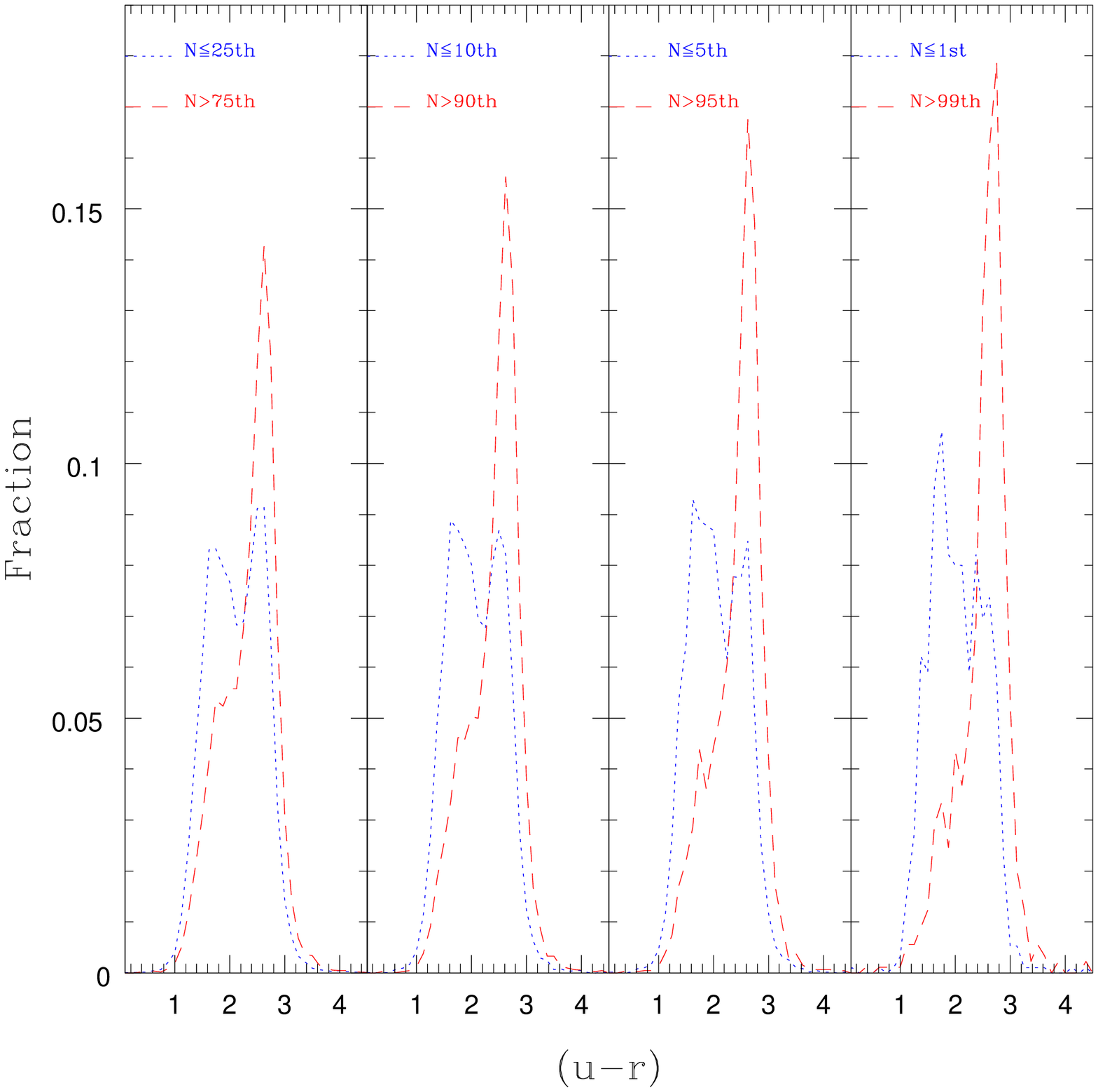}
\caption{($u - r$) color distribution for galaxies in different environments.}
\label{fig:ur_percentile_5Mpc}
\end{center}
\end{figure*}
\begin{figure}[htbp]
\begin{center}
\includegraphics[width=8.5cm]{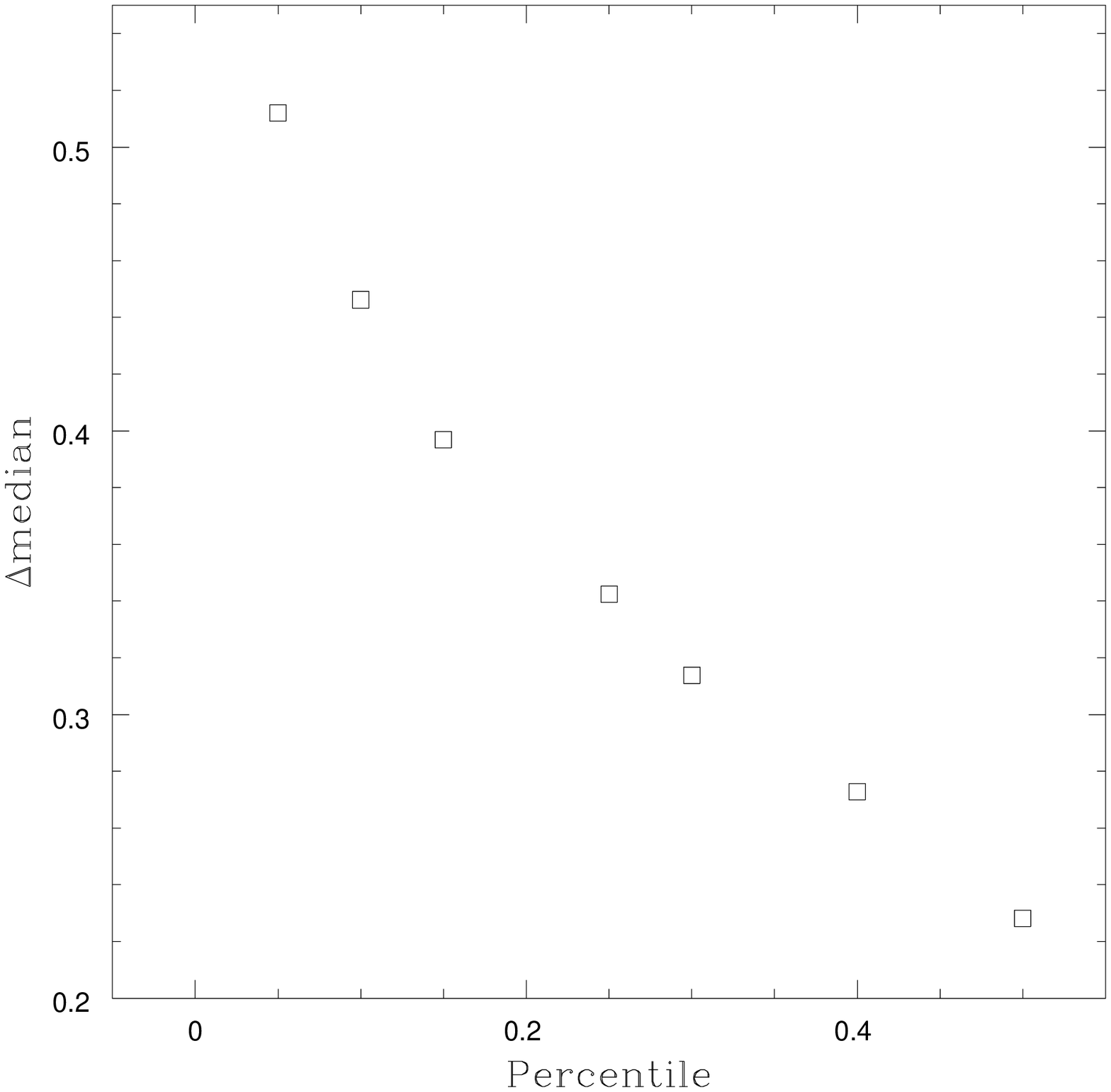}
\caption{The trend of $\Delta$median, defined in the text, as a function of the
percentile.}
\label{fig:ur_percentile_diff}
\end{center}
\end{figure}
\begin{figure*}[htbp]
\begin{center}
\includegraphics[width=12cm]{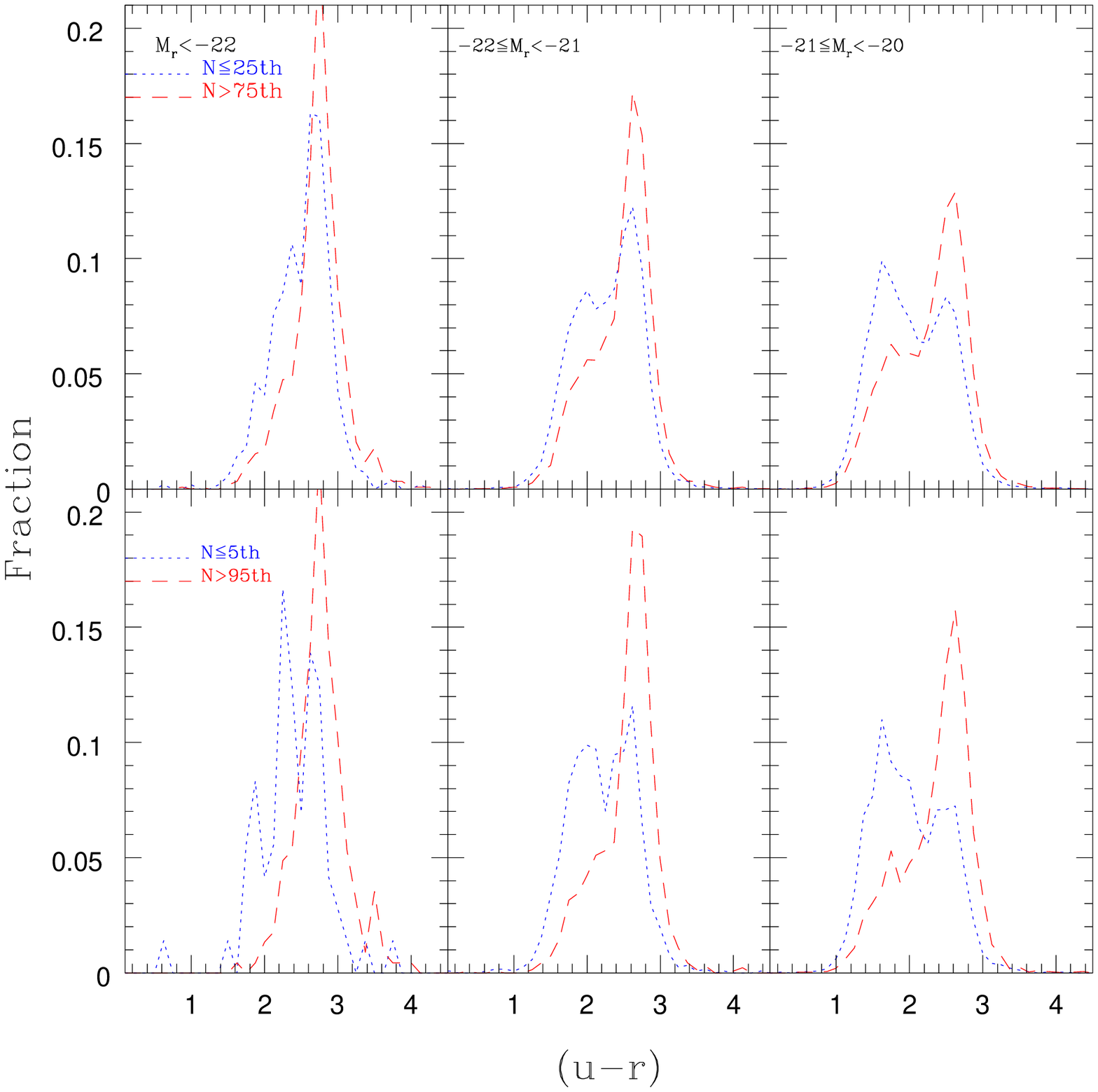}
\caption{Top panels: comparison of the color distribution for galaxies having a
number of neighbours lesser than the 25th percentile with the color
distribution for galaxies having a number of neighbours larger than the 75th
percentile. Bottom panels: comparison of the color distribution for galaxies
having a number of neighbours lower than the 5th percentile with the color
distribution for galaxies having a number of neighbours grater than the 95th
percentile. In both the panels, the luminosity decreases from left to right.}
\label{fig:ur_luminosity}
\end{center}
\end{figure*}

\noindent
It is well known that the environment can affect the colors of galaxies, and
that galaxies in low density environment are generally bluer than galaxies in
clusters \citep{1999AJ....118.2561G}. However, we can not relate the color of
galaxies to their environment because the ($u - r$) color does not correlate with
the local density in a single way.  Then, because the luminosity function of
bright galaxies decreases steeply with increasing luminosity, we investigated
whether the differences between cluster and void galaxies are also a function
of their luminosity, as suggested by \cite{2004ApJ...617...50R}. In
Fig.~\ref{fig:ur1}, we analize the ($u - r$) color distribution for all the
galaxies in our sample (left panel), for galaxies in different environments
(central panel), and for two different ranges of luminosity (right panel). To
this last aim, we adopted the median value of the luminosity distribution to
define two different ranges of luminosities: $M_r$ fainter and brighter than
$M_r = -20.7$ . Galaxies in underdense environments (defined as those having
$N<4$ neighbours within $5$ Mpc, where $N=4$ corresponds to the 10th
percentile) and galaxies fainter than  $M_r$ = -20.7 have a similar, bimodal,
color distribution. The second peak is redder and coinciding with the peak of
the color distribution for galaxies in dense environments ($N>59$ neighbours,
corresponding to the $90^{th}$ percentile) and brighter than $M_r = -20.7$.
From this first analysis, it is evident that, in general, galaxies in
underdense environments are bluer and fainter than galaxies in dense
environments.\\
In Fig.~\ref{fig:ur_percentile_5Mpc}, differences between over- and underdense
regions, as characterized by the number of neighbours, are increasing from left
to right. For instance, in the rightmost panels, the  distributions are
referred to galaxies having more than 127 or less than 1 neighbours within $5$
Mpc, corresponding to the $1^{st}$ and $99^{th}$ percentiles of the global
distribution, respectively. In this way we are able to compare environments
that are more and more extreme. These plots confirm that in underdense regions
galaxies tend to be bluer than cluster galaxies, and the transition from over-
to under-dense regions is smooth.  In Fig.~\ref{fig:ur_luminosity}, the galaxy
populations for different environment are compared, using progressively fainter
galaxies, from left to right. Galaxies in underdense environment ($N_{neigh}\le
25th$ or $N_{neigh}\le 5th$) are bluer at fainter luminosities and the
transition is smooth: this result is confirmed by a KS test and it is also in
accordance with \citet{2004ApJ...617...50R}. Then, we conclude that both
luminosity and density act to discriminate galaxies in underdense environments
from galaxies in the field or in clusters.\\ 
\noindent
In the following, we will make these statements more quantitative.\\
In order to quantify the difference between over- and underdense regions, we
compute a $\Delta median$ value of ($u - r$), defined as the difference of the
median value ($u - r$) between the distributions of two opposite percentile. We
define as opposite percentile the value 1-p, being p  the value of the
percentile. For example, for the $20^{th}$ percentile the opposite percentile
is the $80^{th}$ so that, for this example, $\Delta median_{20th}$ will be: 
\begin{equation}
\Delta median_{20th}=(u-r)_{N<20th}-(u-r)_{N>80th}
\end{equation}
 \noindent
where $(u - r)$ is the median value of the distribution.\\
The plot in Fig.~\ref{fig:ur_percentile_diff} shows a clear trend in the
variation of the previous quantity with the color. It is interesting to observe
that the variation is a smooth function of the difference in percentiles: we do
not observe a very steep change of the slope for small values of the
percentile, i.e., for very different environments. This result is a further
hint at the fact that there is not a ''void galaxy population'' with
radically distinct properties, as suggested by Peebles (2001) 
for a standard CDM model.
\section{($u - r$) color distribution: dependence on morphology}  
\begin{figure}[htbp]
\begin{center}
\includegraphics[width=8.5cm]{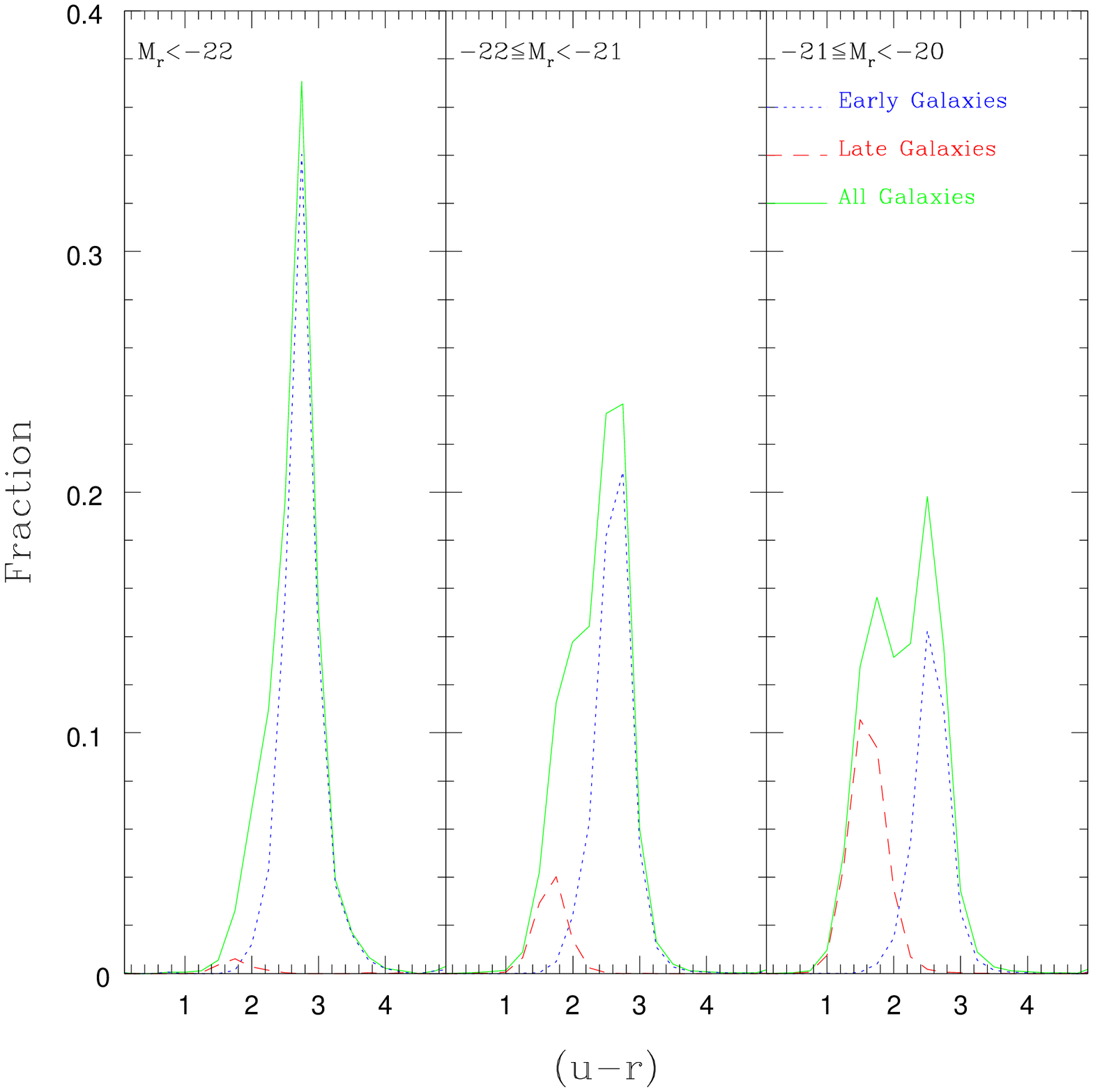}
\caption{($u - r$) color distribution for galaxies having different
luminosities and different morphological type.}
\label{fig:ur_morph_luminosity}
\end{center}
\end{figure}

In the previous section, we studied the dependence of the ($u - r$) color
distribution for the galaxies on both the environment and the luminosity,
without to consider any dependence on their morphological type. But it is well
known that the morphological type of galaxies correlates with density
\citep{1980ApJ...236..351D} and color \citep{2001AJ....122.1861S}, as well as
there is also a correlation between density, luminosity and color 
\citep{2003ApJ...585L...5H}.  We also found that galaxies in underdense
environments are bluer than galaxies in dense environment, in accordance with
\cite{2004AIPC..743..106B}, but it is unclear if this is an induced effect
driven by some other relation, e.g. the morphology-density relation
\citep{1980ApJ...236..351D}. Because the value ($u - r$) = 2.22 can be used to
separate early- from late-type galaxies \citep{2001AJ....122.1861S} and because
a clear bimodality is observed in the ($u - r$) color distribution, in this
section we will investigate the existence of a possible correlation between the
findings in sect. 4 and the morphology. In particular, in order to relate the
morphology to the environment, we compare the ($u - r$) color distribution for
early- and late-type galaxies with the color distribution of the total sample
of galaxies.\\
\begin{table}
\begin{center}
\caption{Number of galaxies in the plots of Fig.~\ref{fig:ur_morph_percentile}}
\label{tab:ur_morph_percentile}
\begin{tabular}{||c|r|r|r||}
\hline
                    & All       & Early (\%) & Late (\%)\\
\hline
\hline
$0\le N<4   $       & 7205      & 31.3       & 31.7      \\
\hline 
$4\le N<7   $       & 8402      & 36.2       & 28.4      \\
\hline
$7\le N<11  $       & 17490     & 38.0       & 26.7      \\
\hline 
$11\le N<18 $       & 17393     & 41.1       & 24.1      \\
\hline
$18\le N<30 $       & 18608     & 45.2       & 21.1      \\
\hline 
$30\le N<41 $       & 9595      & 46.8       & 18.8      \\
\hline
$41\le N<59 $       & 9215      & 50.4       & 16.6      \\
\hline 
$59\le N<127$       & 8470      & 55.1       & 14.0      \\
\hline
\end{tabular}
\end{center}
\end{table}
\begin{figure*}[htbp]
\begin{center}
\includegraphics[width=14cm]{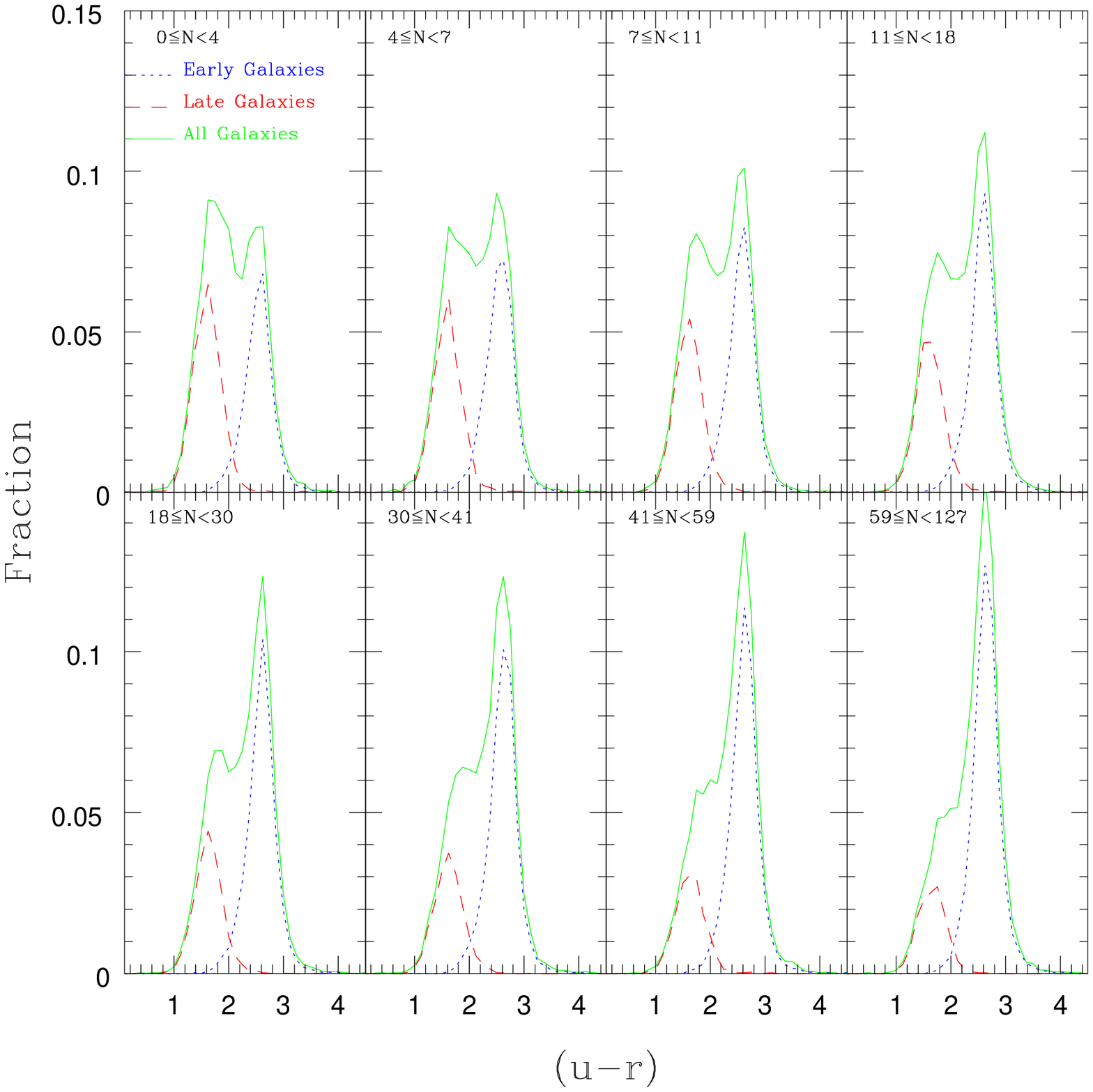}
\caption{$(u - r)$ color distribution for different environment ($N$ is the
number of neighbours within 5 Mpc) and for different morphological types.}
\label{fig:ur_morph_percentile}
\end{center}
\end{figure*}
\noindent
From Fig.~\ref{fig:ur_morph_percentile} we see that the bimodal shape of the
($u - r$) color distribution is a function of both the environment and the
morphological type, becoming redder with increasing the density of the
environments, as well as the fraction of early-type galaxies: in poor systems
(i.e. $0 \le N < 4$) the fractions of early- and late-type galaxies are similar;
with increasing the number of neighbours the fraction of early-type galaxies
also increases and the bimodality almost disappears (i.e. $59 \le N < 127$). 
The numbers $N$ of galaxies used in Fig.~\ref{fig:ur_morph_percentile} are not
casual, but they rather correspond to the percentiles of the total distribution
(Tab.~\ref{tab:percentile}).\\
\begin{table}
\begin{center}
\caption{Number of galaxies in the plots of Fig.~\ref{fig:ur_morph_luminosity}}
\label{tab:ur_morph_luminosity}
\begin{tabular}{||c|r|r|r||}
\hline
                          & All      & Early (\%) & Late (\%) \\
\hline
\hline
$ M_{r}<-22             $ & 3617     & 75.3       & 1.5       \\
\hline 
$ -22 \le M_{r}<-21     $ & 27815    & 55.1       & 9.3       \\
\hline
$ -21 \le M_{r} \le -20 $ & 60134    & 36.0       & 29.8      \\
\hline

\end{tabular}
\end{center}
\end{table}
In the previous section it has also been observed that the bimodal ($u - r$)
color distribution depends on both the environment and the luminosity. In fact,
the fraction of blue galaxies decreases becoming more and more red with
increasing of both the environmental density and the luminosity
(Fig.~\ref{fig:ur_luminosity}), in accordance with \citet{2004AIPC..743..106B}
and \citet{2005ApJ...632...49M}. Then in order to relate this result to the
morphological type, we plot the ($u - r$) color distribution for different ranges
of luminosities and morphologies (Fig.~\ref{fig:ur_morph_luminosity}). We find
that the fraction of early-type galaxies decreases, while the fraction of
late-type galaxies increases in samples which are over fainter, then probing
the environmental dependence on the luminosity of the ($u - r$) color
distribution for early- and late-type galaxies.\\
\begin{figure}[htbp]
\begin{center}
\includegraphics[width=8.5cm]{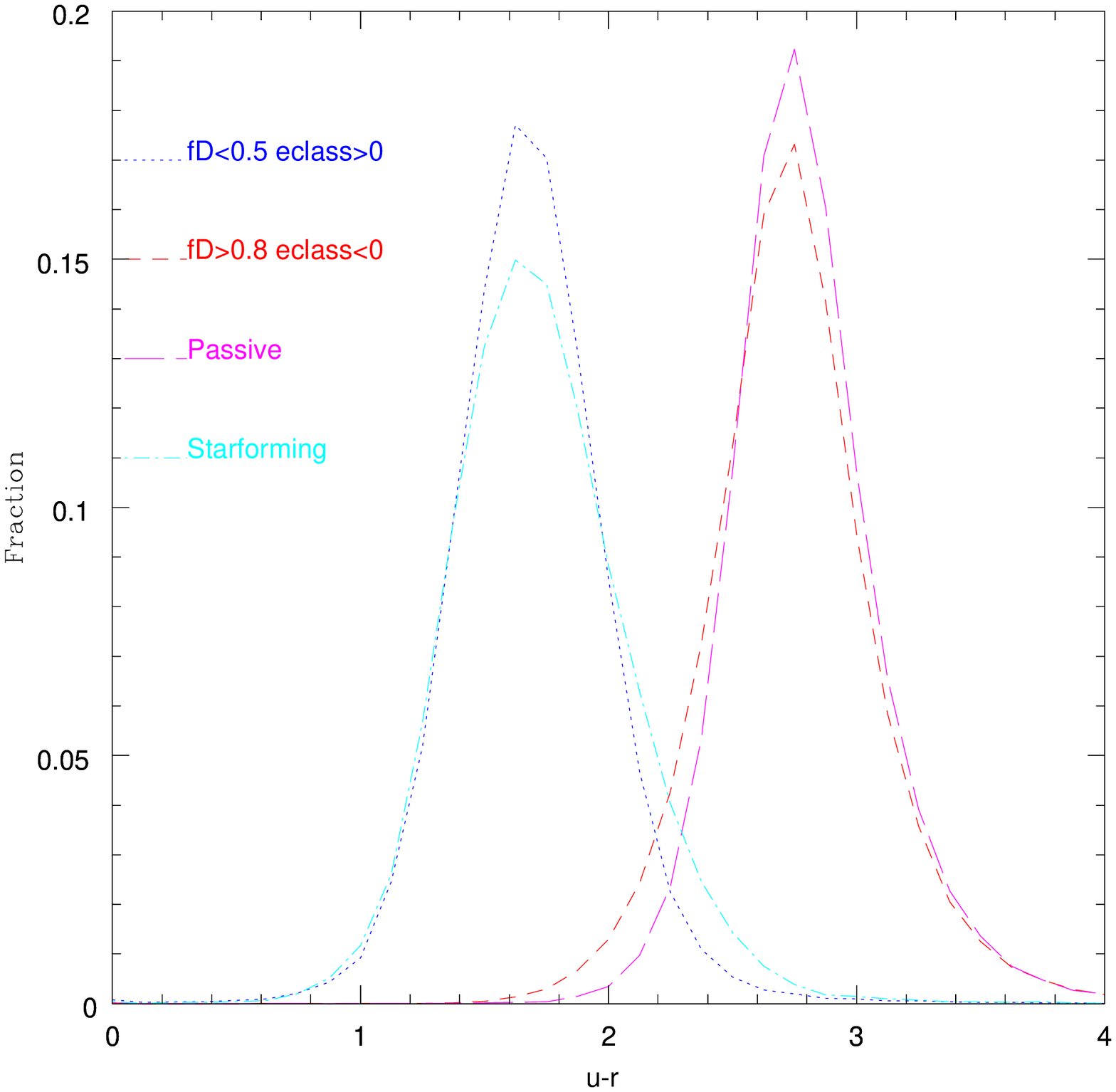}
\caption{The ($u - r$) color distribution for early- and late-type galaxies 
spectroscopically selected by using the presence of emission lines in the 
spectra  and the $FracDev_{r}$ and $Eclass$ parameters as in 
\citet{2005AJ....129...61B}. Both the distributions are in good agreement.}
\label{fig:ur_morphology}
\end{center}
\end{figure}
Finally, we condider the ($u - r$) color distribution for galaxies having
different morphological type and in different bins of magnitudes. Galaxies with
$-21 \le M_r < -20$ have a similar fraction of early (36.0\%) and late (29.8\%)
type galaxies; for $-22 \le M_r < -21$ these fractions are 55.1\% (early) and
9.3\% (late), while for $M_r < -22$ the fraction of late-type galaxies is
negligible (1.5\%) compared to the fraction of early-type (75.3\%). In
general, we have a similar trend (Fig.~\ref{fig:ur_morph_luminosity}): the
fraction of late-type galaxies decreases with increasing of both the
environmental density and luminosity, while the fraction of early-type galaxies
increases, then confirming the previous findings concerning the environmental
dependence on the luminosity of the ($u - r$) color distribution for early- and
late-type galaxies. 
Moreover, the color distributions for early-type/PGs and
late-type/SFGs galaxies defined as described in sec. 3., are similar
(Fig.~\ref{fig:ur_morphology}), so that we can relate the environmental
properties of late-type galaxies not only to the {\it morphology-density}
relation \citep{1980ApJ...236..351D}, but also to their star-forming
activity.\\
\section{Discussion}
By definition, there are few galaxies in voids, and only the availability of
large samples can allow to build statistically significant catalogues of
galaxies in such underdense regions. These conditions are met by the sample we
use in this paper, extracted from the SDSS-DR4.\\
\noindent
Studying the galaxy
properties in extremely underdense environments is an interesting test for
cosmological and galaxy formation models. In standard CDM models, underdense
regions should be populated with sub-L$^{\ast}$ galaxies
\citep{1986ApJ...303...39D}. 
Many of these "void dwarfs" were successfully identified in
previous surveys. For example \citet{1995A&A...301..329L,
1996A&A...314....1L} found that their density decreases
towards the center of the voids themselves. 
Recent surveys like the SDSS show the presence of both 
isolated galaxies and filaments thereof within voids
\citep{2003A&A...405..425E}, a circumstance which is also 
the result of numerical simulations \citep{2002MNRAS.332....7A,2003MNRAS.344..715G}. 
Because these coherent structures are difficult to
identify in galaxy surveys, recently many effort have benn carried out
to detect filaments \citep{2005MNRAS.358..256P}.\\ 
\citet{2001ApJ...557..495P} notices that the existence of
a particular class of void-galaxies marked by a discontinuity in the observed
properties of galaxy populations, from over- to underdense regions,
would be a distinctive feature of galaxy formation in CDM models. On the
contrary, our results (e.g., Figure~\ref{fig:ur_luminosity}) point out a smooth
transition in colors and other properties with average density. This result is
also confirmed when we consider the difference between truly void-galaxies and
''wall'' galaxies, then suggesting that does not exist a "pure" void-galaxy
population, distinguished from the average  galaxy population.\\
However, the only study of galaxy colors as a function of the environment 
gives a partial view. In fact, the color of galaxies in underdense environments
has a different distribution from that of galaxies in dense environment (red),
as shown in Fig.~\ref{fig:ur1}. On the contrary, in underdense environments,
the distribution of the galaxy populations without taking into account the
morphological type is not equally skewed towards blue colors, confirming the
findings by \citet{2004ApJ...615L.101B}. Galaxies become bluer only if we
consider a very underdense environment.\\
When we consider galaxies fainter than $M_r=-22.0$, their color becomes more
and more blue in environments that are more and more underdense
($N_{neigh}<25^{th}$, $N_{neigh}<10^{th}$, $N_{neigh}<5^{th}$, $N_{neigh}<1^{st}$). 
A similar trend (shifted to the red) is observed for galaxies in environments
more and more dense ($N_{neigh}>75^{th}$, $N_{neigh}>90^{th}$, $N_{neigh}>95^{th}$,
$N_{neigh}>99th$), (Fig.~\ref{fig:ur_percentile_5Mpc}). This trend becomes more
and more evident when we consider galaxies more and more faint. We find that in
dense environments, galaxies fainter than $M_r=-21$ are redder than galaxies
in underdense environments (Fig.~\ref{fig:ur_luminosity}), confirming that the
color distribution of galaxies is also strongly dependent on the absolute
magnitude. Then, we can assert that the ($u - r$) color distribution of galaxies
is both related to the environment and to the luminosity of the galaxies.

These results are in agreement with \citet{2005MNRAS.356.1155C} 
and \citet{2005A&A...439...45E} who showed that luminosity function parameters
and brightest luminosities of galaxies are smooth functions of the density 
of the environment.\\

When we take into account the morphological type of galaxies, we find that the
fraction of late-type is bluer and fainter than the fraction of early-type
galaxies, and that their fraction decreases from low ($0\le N<4$) to dense
environment ($59\le N<127$), while the early-type galaxies fraction increases
(Fig.~\ref{fig:ur_morph_percentile}). In particular, in the range of
luminosities brighter than $M_r=-22.0$, we have an excess of early-type
(75.3\%)  respect to late-type (1.5\%) galaxies, while for low luminosities
($M_r>-21$) the fraction of late-type galaxies increases (29.8\%) compared to
fraction of early-type (36.0\%) (Fig.~\ref{fig:ur_morph_luminosity} and
Tab.~\ref{tab:ur_morph_luminosity}). This result is in accordance with the {\it
morphology-density} relation \citep{1980ApJ...236..351D}, confirming that, on
average, early-type galaxies are redder, brighter, and in denser environments
than late-type galaxies.\\
It is important to notice that the ($u - r$) color distributions for early-type
and late-type galaxies, defined through the SDSS Eclass and FracDev parameters,
are similar to the color distribution for PGs and SFGs, respectively
(Fig.~\ref{fig:ur_morphology}). Then, the environmental properties of late-type
galaxies can be related to the {\it morphology-density} relation
\citep{1980ApJ...236..351D} not only as a consequence of their morphological
type, but also for their {\it star-forming} activity. This latter point is not
completely unexpected. In fact, in the sample of galaxies used by
\citet{2000AJ....119...32G,1999AJ....118.2561G}, there are a lot of galaxies in
voids that are found in pairs or small groups. Thus, enhanced stellar formation
due to tidal effects is likely to be detected among void galaxies.\\
\noindent
In a recent paper, \citet{2006astro.ph..5703P} analysed the properties of voids 
galaxies in the SDSS-DR4, founding some results similar to ours, i.e., a bimodal 
distribution of the ($u - r$) color index and a general trend for galaxies in voids 
to be bluer than in clusters: this latter result is strongly dependent on the 
absolute magnitude, as we demonstrate in this paper. In particular, they also 
found no difference between void and field galaxies if only galaxies in a restricted 
absolute magnitude range are analysed.


\section{Conclusions}
In this paper we analized the environmental dependence of galaxy population
properties, looking for those features which could hint at the presence of a
"void" galaxy population. The photometric properties we used in our analysis
can be regarded as complementary respect to those used by other authors 
\citep{2004ApJ...617...50R, 2005ApJ...620..618H}. In particular, our density
estimator allows us to exploit redshift informations, then  finding galaxies in
very underdense regions.\\
We can summarize our findings in a few points:
\begin{enumerate}
\item Luminosity is a sensitive parameter to characterize the environmental
properties of galaxies.
\item On average, a very underdense environment ($N_{neigh}\le4$) is populated
by galaxies which are bluer than galaxies that are in a very dense enviroments,
($N_{neigh}>79$, see Fig.~\ref{fig:ur1}).
\item On average, faint galaxies ($M_{r}>-21.0$) are bluer in very underdense
environments than in dense environments and  the transition from blue to red,
moving from under- to overdense environments, is more pronounced for faint
galaxies ($M_{r}>-21.0$, Fig.~\ref{fig:ur_luminosity}).
\item Although the transition from voids to clusters is more pronounced for
faint galaxies, it does not seem to be a discontinuity
(Fig.~\ref{fig:ur_percentile_diff}).
\item Changes in the ($u - r$) color distribution are related to the environment,
to the luminosity, and to the morphology of the galaxies. In fact, on average,
fainter galaxies ($M_r>-21.0$) are bluer, late-type and in underdense regions
(voids)  than brighter galaxies ($M_r<-21.0$), which are redder, early-type and
in overdense regions (clusters).  
\item Galaxies classified as early-type/late-type and PGs/SFGs, according to
their photometric/spectroscopic parameters and to their class of star-forming
activity, have a similar ($u - r$) color distribution
(Fig.~\ref{fig:ur_morphology}). Then, the environmental properties of late-type
galaxies can be related to the {\it morphology-density relation}
\citep{1980ApJ...236..351D} not only as a consequence of their morphological
type, but also as a consequence of their star-forming activity. 
\item We don't find any sudden transition in the properties of "void" galaxies,
respect to cluster galaxies, as suggested by Peebles (2001). On the contrary,
our results show a continuity in the properties of the galaxies, from voids to
clusters.
\end{enumerate}

\begin{acknowledgements}
We are grateful to Chris C. Haines for having carefully read the manuscript and
for the useful conversations and comments.\\
We wish to thank the referee, Jaan Einasto, for the constructive comments which
improved the content of the paper.\\ 
G.S. and A.R. thank the EC and the MIUR for having partially supported this
work (EC contracts HPRN-CT-2002-00316 - SISCO network, and MIUR-COFIN-2004
n.2004020323\_001).\\
V.A. gratefully acknowledges the support from the European Commission
(Marie Curie ''Transfer of Knowledge'' Project COSMOCT, contract
MC-TOK-002995.)\\
Funding for the SDSS and SDSS-II has been provided by the Alfred P. Sloan
Foundation, the Participating Institutions, the National Science Foundation,
the U.S. Department of Energy, the National Aeronautics and Space
Administration, the Japanese Monbukagakusho, the Max Planck Society, and the
Higher Education Funding Council for England. The SDSS Web Site is
http://www.sdss.org/.\\ The SDSS is managed by the Astrophysical Research
Consortium for the Participating Institutions. The Participating Institutions
are the American Museum of Natural History, Astrophysical Institute Potsdam,
University of Basel, Cambridge University, Case Western Reserve University,
University of Chicago, Drexel University, Fermilab, the Institute for Advanced
Study, the Japan Participation Group, Johns Hopkins University, the Joint
Institute for Nuclear Astrophysics, the Kavli Institute for Particle
Astrophysics and Cosmology, the Korean Scientist Group, the Chinese Academy of
Sciences (LAMOST), Los Alamos National Laboratory, the Max-Planck-Institute for
Astronomy (MPA), the Max-Planck-Institute for Astrophysics (MPIA), New Mexico
State University, Ohio State University, University of Pittsburgh, University
of Portsmouth, Princeton University, the United States Naval Observatory, and
the University of Washington.
\end{acknowledgements}

\bibliographystyle{/home/van/tex/A_and_A/6.0/bibtex/aa}
\bibliography{biblio}


\appendix
\section{THE ENVIRONMENT SMOOTHING SCALE}

\begin{figure*}[htbp]
\begin{center}
\includegraphics[width=14cm]{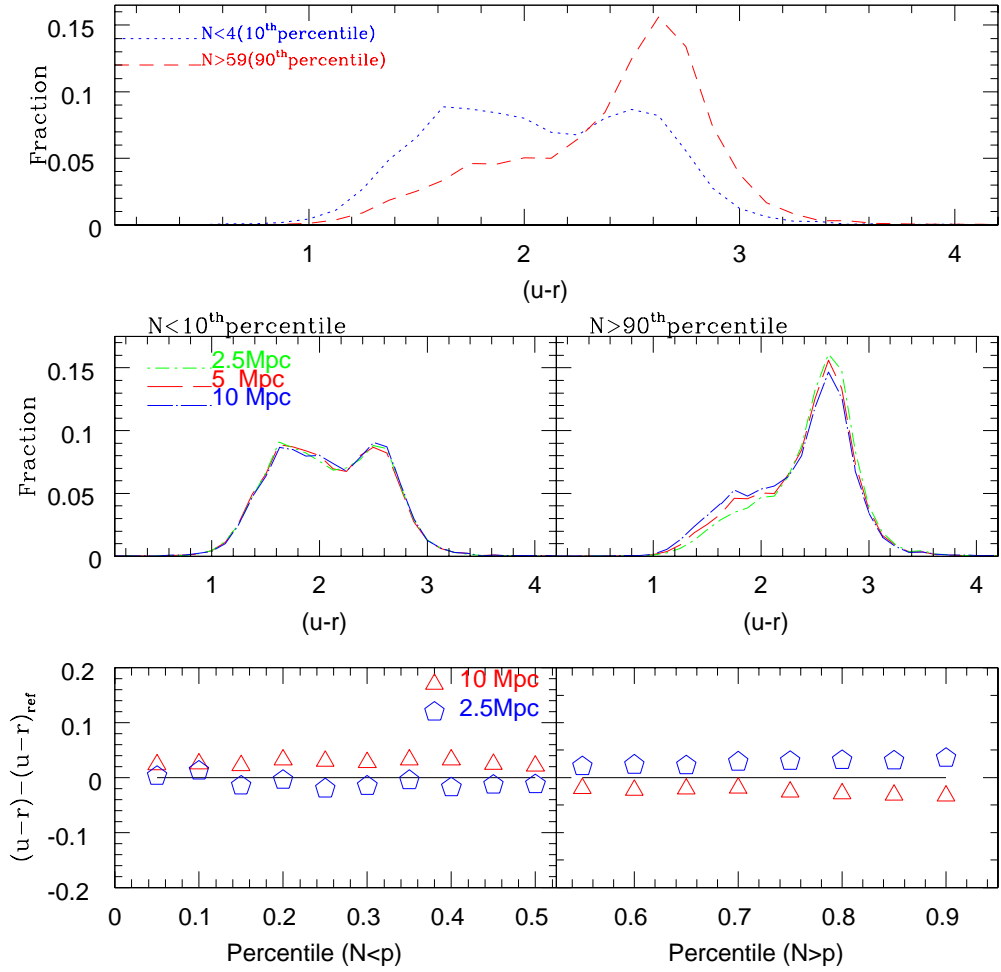}
\caption{Top panel: the ($u - r$) color distribution for a smoothing scale 
value of 5 Mpc (the reference value), in the underdense and overdense 
environments. 
Middle panel: as in the top panel, for three different values of smoothing 
scale: 2.5 Mpc, 5 Mpc, and 10 Mpc (left panel:underdense environments; 
right panel:overdense environments). 
Bottom panel: differences in the ($u - r$) color distribution respect to the 
reference values, for different environments (left panel:underdense; 
right panel:overdense).}
\label{fig:ref}
\end{center}
\end{figure*}

In this appendix, we will examine how changing the environment smoothing 
scale with which we define the local galaxy density affects our results.
To this aim, we compare the ($u - r$) color distribution for the galaxies in 
our sample in different environments, using three different smoothing 
scale: 2.5 Mpc, 5 Mpc (the reference value adopted in this paper), and 10 Mpc. 
These values are similar to those adopted by  \citet{2005A&A...439...45E}. 
We define the underdense environment $N < 4$ ($10^{th}$ percentile) 
and the overdense environment for $N > 59$ ($90^{th}$ percentile), as in the paper. 
In the top panel of Fig.~\ref{fig:ref} we report the central panel of Fig.~\ref{fig:ur1}. 
In the middle panel we compare the ($u - r$) color distribution for underdense (left) 
and overdense (right) environments, using three different smoothing scale: 2.5 Mpc, 
5 Mpc (the reference value), and 10 Mpc. Finally, in the bottom panel we quantify 
the differences in the previous plots, that are the differences of the median ($u - r$) 
color distribution for 2.5 and 10 Mpc respect to the reference value of of the ($u - r$) 
color distribution for 5 Mpc adopted in the paper. These differences are plotted both 
for the underdense (left panel) and overdense (right panel) environments, with values 
$(u-r) - (u-r)_{ref} < 0.03$.

The previous analysis demonstrates that the adopted value of smoothing scale (5 Mpc) 
is a robust representation and a good probe of both the underdense and overdense 
regions of the survey volume.


\end{document}